\pdfoutput=1
\documentclass[twocolumn,showpacs,amssymb,floatfix,aps]{revtex4}

\usepackage{graphicx}
\usepackage{dcolumn}
\usepackage{bm}
\usepackage{epsfig}

\begin{document}
CERN-PH-TH/2010-201
\preprint{CERN-PH-TH/2010-201}
\title{
QED is not endangered by the proton's size}
\author{A. De R\'ujula${}^{a,b,c}$}
\affiliation{  \vspace{3mm}
${}^a$ Instituto de F\'isica Te\'orica, Univ. Aut\'onoma de Madrid, Madrid, and 
CIEMAT, Madrid, Spain,\\
${}^b$ Physics Dept., Boston University, Boston, MA 02215,\\
${}^c$Physics Department, CERN, CH 1211 Geneva 23, Switzerland}

\date{\today}

\begin{abstract}

Pohl {\it et al.}~have reported a very precise
measurement of the Lamb-shift in muonic Hydrogen \cite{Pohl},
from which they infer the radius characterizing the proton's charge distribution. The result is 5
standard deviations away from the one of the CODATA
compilation of physical constants. 
This has been interpreted \cite{Pohl} as 
possibly requiring a 4.9 standard-deviation modification of the Rydberg constant, to a new value that would be precise to 3.3 parts in $10^{13}$, as well as putative
evidence for physics beyond the standard model \cite{Flowers}.
I demonstrate that these options are unsubstantiated.
\end{abstract}


\maketitle

\section{Introduction}

The issue is extremely simple. The discrepancy quoted in the abstract is
between results which do not depend on a specific model of the proton's
form factor and results, by Pohl et al., which do \cite{PohlAddendum}.
The conclusion is not that the experiments or the theory are wrong,
but that the model (the customary dipole form factor) is inadequate at
the level of precision demanded by the data. The experiments and
QED are right, the dipole is wrong. 

The conclusion of the previous paragraph is the expected one.
The dipole form factor
is but a rough description of higher-energy data and
is unacceptable on grounds of the analyticity requirements stemming
from causality and the locality of fundamental interactions. 

Moreover, any
simple one-parameter description of the proton's non-relativistic
Sacks form factor, $G_E(-{\mathbf q}^2)$ in terms of only one mass 
parameter is inaccurate: the proton is not so simple.
More precisely, the proton's relativistic form factor, $G_E({q}^2)$, is expected,
in the timelike domain $q^2\!\geq\! 0$, to have a complex structure,
with a first cut starting at $q^2\!=\!4\,m_\pi^2$ and a plethora of branch 
cuts and complex resonant poles thereafter \cite{DispersionRelations}.

The same is true of the charge distribution, $\rho_p(r)$, 
the Fourier transform of $G_E(-{\mathbf q}^2)$. 
Even most naively, $\rho_p(r)$ is expected to have 
a ``core" and a ``pion cloud" \cite{PionCloud}, corresponding to a minimum of two 
length parameters.

\section{In detail}

Let $\ell$ denote an electron or a $\mu^-$.
The leading proton-size correction to the energy levels of an $\ell p$ atom
is
\begin{eqnarray}
\Delta E&=&{2\,\alpha^4\over 3\,n^3}\,m_r^3\,\delta_{l 0}\,\langle r_p^2\rangle 
\nonumber\\
m_r&\equiv&{m_\ell\,m_p\over m_\ell+m_p}
\label{delta2}
\end{eqnarray}
where $\langle r_p^2\rangle$ is the mean square radius of  $\rho_p(r)$.

The charge distribution is 
related to the non-relativistic limit of the 
electric form-factor, $G_E$, by the Fourier transformation 
\begin{equation}
G_E(-{\mathbf q}^2)=\int d^3 r \rho_p({\mathbf r})\, e^{-i\,\vec q\, \vec r}
\label{F}
\end{equation}

 Precise measurements of $\langle r_p^2\rangle$ have two origins. One is
mainly based on the theory \cite{HTheory} and observations  \cite{Hexps} of  
Hydrogen. The 
result, compiled in CODATA \cite{CODATA}, is
\begin{equation}
{\langle r_p^2\rangle}\rm (CODATA)=(0.8768 \pm 0.0069\; \rm f)^2
\label{rCODATA}
\end{equation}
The second type of measurement is based on the theory and observations
\cite{Sick, Sick2} of very low-energy electron-proton scattering. It yields
\begin{equation}
{\langle r_p^2\rangle}(ep)=(0.895 \pm 0.018\; \rm f)^2
\label{rep}
\end{equation}
This result requires a sophisticated data analysis, partly
based on a continued-fraction expansion of $G_E$ \cite{Sick}. 

The two quoted methods of measuring $\langle r_p^2\rangle$
are model-independent, in the sense of not
assuming a particular form of the proton's charge distribution, $\rho_p(r)$.

The plot thickens as one considers the 
Lamb shift $\rm 2P_{3/2}^{F=2}\to 2S_{1/2}^{F=1}$ in the $\mu p$ atom, 
measured \cite{Pohl} to be
\begin{equation}
L_{\rm exp} = 206.2949 \pm 0.0032\;\rm meV.
\label{Lexp}
\end{equation}
In meV units for energy and fermi units for the radii, the predicted value 
\cite{LymanTH}
is of the form
\begin{eqnarray}
L^{\rm th}\left[\langle r_p^2\rangle,\langle r_p^3 \rangle_{(2)}\right]&=&\nonumber\\
209.9779(49)&-&5.2262\, \langle r_p^2\rangle +0.00913 \,\langle r_p^3 \rangle_{(2)}
\label{Lth}
\end{eqnarray}
The first two coefficients are best  estimates of many contributions 
 while the third stems from the $n=2$ value of an addend
  \cite{FriarSick,HTheory}
 \begin{equation}
 \Delta E_3(n)= {\alpha^5\over 3\,n^3}\,m_r^4 \,
 \delta_{l0}\,\langle r_p^3 \rangle_{(2)},
 \label{DeltaE3}
 \end{equation}
 proportional to the third Zemach moment
\begin{equation}
\langle r_p^3 \rangle_{(2)}\equiv\int d^3 r_1 d^3 r_2\,\rho(r_1)\rho(r_2)
\vert {\mathbf r}_1-{\mathbf r}_2\vert^3
\end{equation}

For a single-parameter description of the charge distribution, there is an
explicit relation between $ \langle r_p^3 \rangle_{(2)}$ and $\langle r_p^2\rangle$.
Consider, as an example, a $\rho$-dominated form factor in its narrow-width
non-relativistic limit
\begin{equation}
G_E(q^2)={m_\rho^2\over q^2-m_\rho^2+i\,m_\rho\,\Gamma_\rho}
\to{m_\rho^2\over {\mathbf q}^2+m_\rho^2}
\end{equation}
The corresponding charge distribution is a Yukawian
\begin{equation}
\rho(r)={m_\rho^2\over4 \,\pi\, r}e^{-m_\rho r}  
\end{equation}
Its relevant moments are $\langle r^0\rangle\!=\!1$, 
\begin{equation}
\langle r^2\rangle\!=\!6/m_\rho^2\,,\;\;
\langle r^3\rangle\!=\!24/m_\rho^3\,,\;\;
\langle r^3\rangle_{(2)}\!=\!60/m_\rho^3
\label{polemoments}
\end{equation}
The model-dependent relation is thus
\begin{equation}
\left[\langle r^3\rangle_{(2)}\right]^2= {50\over 3}\, \left[\langle r^2\rangle\right]^3
\label{OnePoleRelation}
\end{equation}

For a dipole form factor
\begin{equation}
G_E(-{\mathbf q}^2)={m_d^4\over ({\mathbf q}^2+m_d^2)^2}
\end{equation}
the charge distribution is an exponential
\begin{equation}
\rho(r)={m_d^3\over 8 \,\pi}\,e^{-m_d r}  
\end{equation}
for which $\langle r^0\rangle=1$, 
\begin{equation}
\langle r^2\rangle\!=\!12/m_d^2\,,\;
\langle r^3\rangle\!=\!60/m_d^3\,,\;
\langle r^3\rangle_{(2)}\!=\!315/(2\,m_d^3)
\label{dipolemoments}
\end{equation}
The model-dependent relation is thus
\begin{equation}
\left[\langle r^3\rangle_{(2)}\right]^2={3675\over 64} \left[\langle r^2\rangle\right]^3
\label{DipoleRelation}
\end{equation}

The ratio of the numbers in 
Eqs.~(\ref{OnePoleRelation},\ref{DipoleRelation})
 is $441/512\!\sim\! 0.86$, showing the difference of relevant moments
between to two form-factor ``models". Even if we took the sixth root of this
number to bring it closer to unity --as experimentalists do with 
$\langle r^2\rangle$ to halve the relative error-- the result would,
at the required great precision, still epitomize  the model-dependence
of the results.

\subsection{A toy model}

The photon propagator in the time-like domain ($q^2\!>\!0$) has 
led to considerable revolutions (e.g.~the discovery and interpretation
of the $J/\Psi$), as well as
interesting challenges, in particular close to its cut at $q^2\!\geq\! 4\,m_\pi^2$.
The modeling of the
electric and magnetic form factors $G_E$ and $G_M$ of protons and neutrons 
in terms of dispersion relations for the photon propagator  
involves, literally, dozens of parameters \cite{DispersionRelations}. 
The form-factor ``toy model" I am going to discuss 
is not intended to compete in accuracy with the dispersive approaches, but only to
elucidate the current discussion.

\begin{figure}[htbp]
\begin{center}
\includegraphics[width=0.40\textwidth]{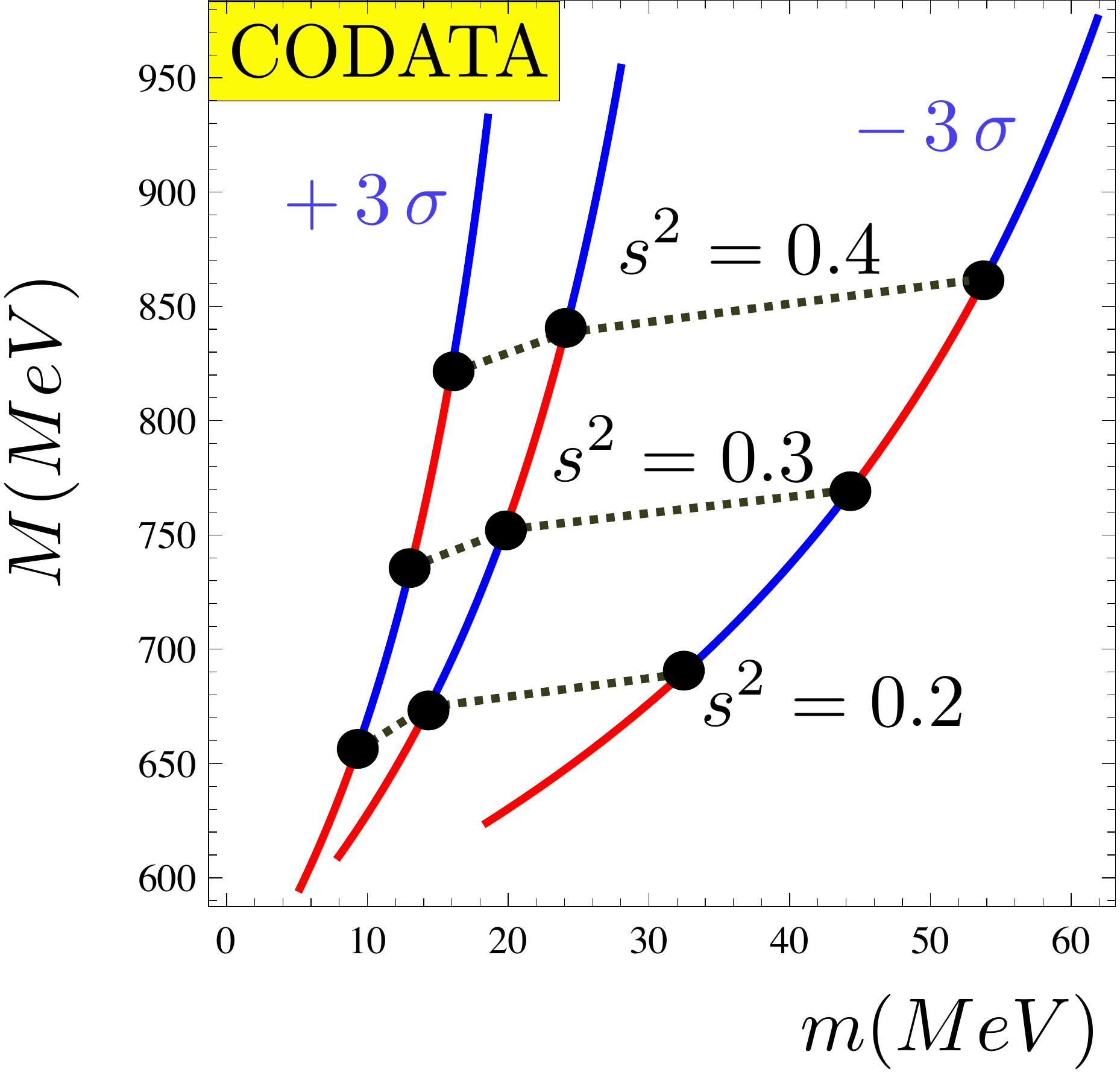}\\
\includegraphics[width=0.40\textwidth]{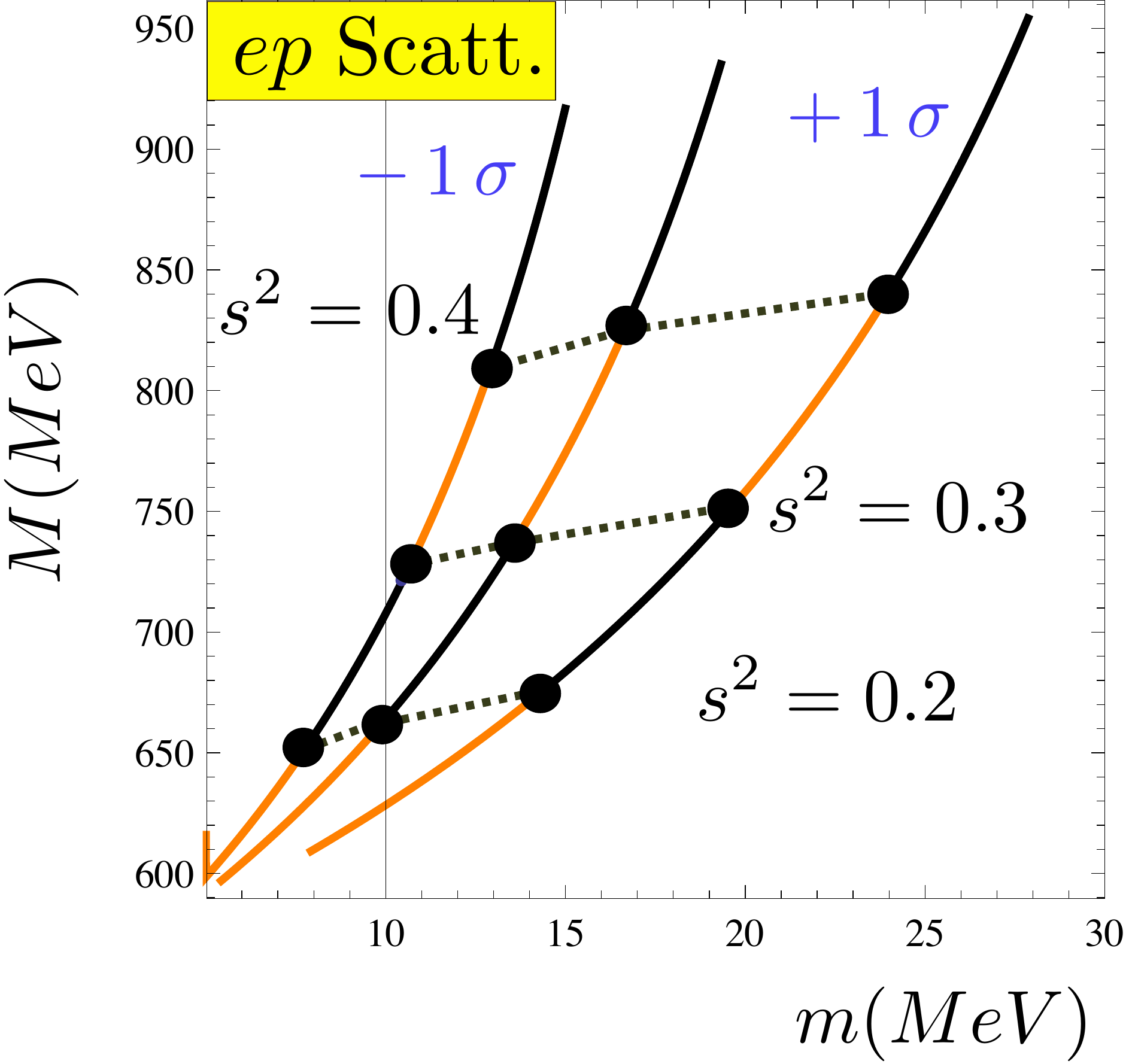}\\
\caption{Parameters $M$ and $m$ for which the toy model is compatible with the data,
with $s^2\!=\!\sin^2(\theta)$ varying along the curves, see 
Eqs.~(\ref{r2comparison},\ref{solCODATA}).
Top: Lyman in the $\mu p$ atom and CODATA,
shown for the central value and a very asymmetric $\pm\, 3\,\sigma$. 
Bottom: CODATA substituted for $e\,p$ scattering, central value and $\pm 1\,\sigma$
(there is no solution for $+\,3\,\sigma$).
  \label{fig:Compatible}}
\end{center}
\end{figure}

In \cite{DispersionRelations}, an accurate description
of the theoretically-calculated $2\pi$ continuum required
products of up to three poles. I parametrize $\rho(r)$ as an interpolation
between the charge densities of a ``$\rho$" single pole and a ``$2\pi$" dipole:
\begin{eqnarray}
 \rho(r)&=&{1\over D}\left[\frac{M^4 e^{-M r}
   \cos ^2(\theta )}{4 \pi  r}+\frac{m^5 e^{-m r} \sin ^2(\theta)}{8 \pi }\right]\nonumber\\
 D&\equiv&  M^2 \cos^2(\theta )+m^2 \sin ^2(\theta )
\end{eqnarray}
   whose two first relevant moments are $\langle r^0\rangle\!=\!1$ and
 \begin{equation}
\langle r^2\rangle\vert_{\rm toy}=   
\frac{6}{m^2 \tan ^2(\theta
   )+M^2}+\frac{12}{m^2+M^2 \cot
   ^2(\theta )}
\end{equation}
To introduce the third Zemach moment, 
let $s\equiv \sin(\theta)$ and $c\equiv \cos(\theta)$.
Then
\begin{eqnarray}
&&   \langle r^3\rangle_{(2)}\vert_{\rm toy}=
\frac{3 \left[5 m M \left(8 c^4
   M+21 m s^4\right)+16 c^2 H
   s^2\right]}{2 m M \left(c^2
   M^2+m^2 s^2\right)^2}\nonumber\\
&&H\equiv\\
&&\frac{ 2 m^5+4 m^4 M+6 m^3
   M^2+8 m^2 M^3+10 m M^4+5
   M^5}{(m+M)^2}\nonumber
\label{Zemach3TwoY}
   \end{eqnarray}
We can now check the compatibility of the CODATA result of Eq.~(\ref{rCODATA})
with the Lamb shift result of Eq.~(\ref{Lexp}) in the following way. Solve the two
equations
\begin{equation}
{\langle r_p^2\rangle}\rm (CODATA)=\langle r^2\rangle\vert_{\rm toy}
\label{r2comparison}
\end{equation}
\begin{equation}
L_{\rm exp} =
L^{\rm th}\left[\langle r^2\rangle,\langle r^3 \rangle_{(2)}\right]\vert_{\rm toy}
\label{solCODATA}
\end{equation}
in $M$ and $m$ for fixed mixing (fixed $s^2$). The results are shown in the top
Fig.~\ref{fig:Compatible}, while those of a similar exercise with 
${\langle r_p^2\rangle}\rm (CODATA)$ substituted by
${\langle r_p^2\rangle}(ep)$ in Eq.~(\ref{r2comparison}) are shown in the bottom
panel.

From these figures we can draw three conclusions:\\
1) The system of Eqs.~(\ref{r2comparison},\ref{solCODATA})
is soluble only for $s^2\!\geq \!0.1$. 
A single-pole or single-dipole are excluded, as expected.\\ 
2) The extracted  $M$ and $m$ are not unreasonable. $M$
turns out to be of ${\cal{O}}(m_\rho)$, while $m$, which corresponds
to a dipole parametrization of ``everything but the $\rho$ pole" is not
a good enough simplification, a result with $m>2\,m_\pi$ would have been
nicer. Yukawa intuited pions in a very similar manner, but only one at a time.\\
3) All experimental results are compatible.

\section{Conclusions}

We face a choice between the following conclusions:
\begin{itemize}
\item{}The experimental results are not right.
\item{}The relevant QED calculations are incorrect.
\item{}There is, at extremely low energies and at the level of accuracy of the $\ell p$-atom
experiments, ``physics beyond the standard model".
\item{}A single-dipole form factor is not adequate to the analysis of precise 
low-energy data. 
\end{itemize}
I have argued that the last choice is the most compelling.

The theoretical and experimental results I have quoted
momentarily culminate 125 years
of progress in the understanding of Hydrogen and its muonic sibling
(I am setting $t=0$ at the discovery date \cite{Balmer} of his famous ``series"
by the Swiss physicist Johann Jakob Balmer).

The combination of the very impressive results in 
Eqs.~(\ref{rCODATA},\ref{Lexp},\ref{Lth}) yields a value:
 \begin{equation} 
\left[ \langle r_p^3 \rangle_{(2)} \right]^{1/3}=3.32\pm0.22\;\rm f
\label{finalresult}
\end{equation}
with the error dominated by the CODATA uncertainty
on $\langle r_p^2\rangle$.
The value in Eq.~(\ref{finalresult}) looks {\it incredibly}
 large at first, but it is not so unexpected:
 
 The result Eq.~(\ref{finalresult}) is $\rho_p(r)$-independent; to
 be treated with due respect. Right after offering
 excuses, I shall break this rule.
The third Zemach moment is very sensitive to the long-distance part of
$\rho(r)$, compare it to the $r^3$ moments of 
Eqs.~(\ref{polemoments},\ref{dipolemoments}).
 Suppose that $\rho(r)$ has a ``core" and a ``tail"
 contributing 50-50 to the proton's charge,
and that the tail's $G_E({\mathbf q})$ is described by a dipole. 
To what scale, $m$, to does this tail correspond?
The value of $\langle r_p^3 \rangle_{(2)}$
is 1/2 of the one in Eq.~(\ref{dipolemoments}). Equate it to
Eq.~(\ref{finalresult}) to obtain
 $m\simeq 245$ MeV, tantalizingly close to the threshold of the proton
form factor's cut at $2\,m_{\pi^\pm}\simeq 278$ MeV.

\begin{figure}[hbt!]
\hspace{-.4cm}
\includegraphics[width=0.5\textwidth]{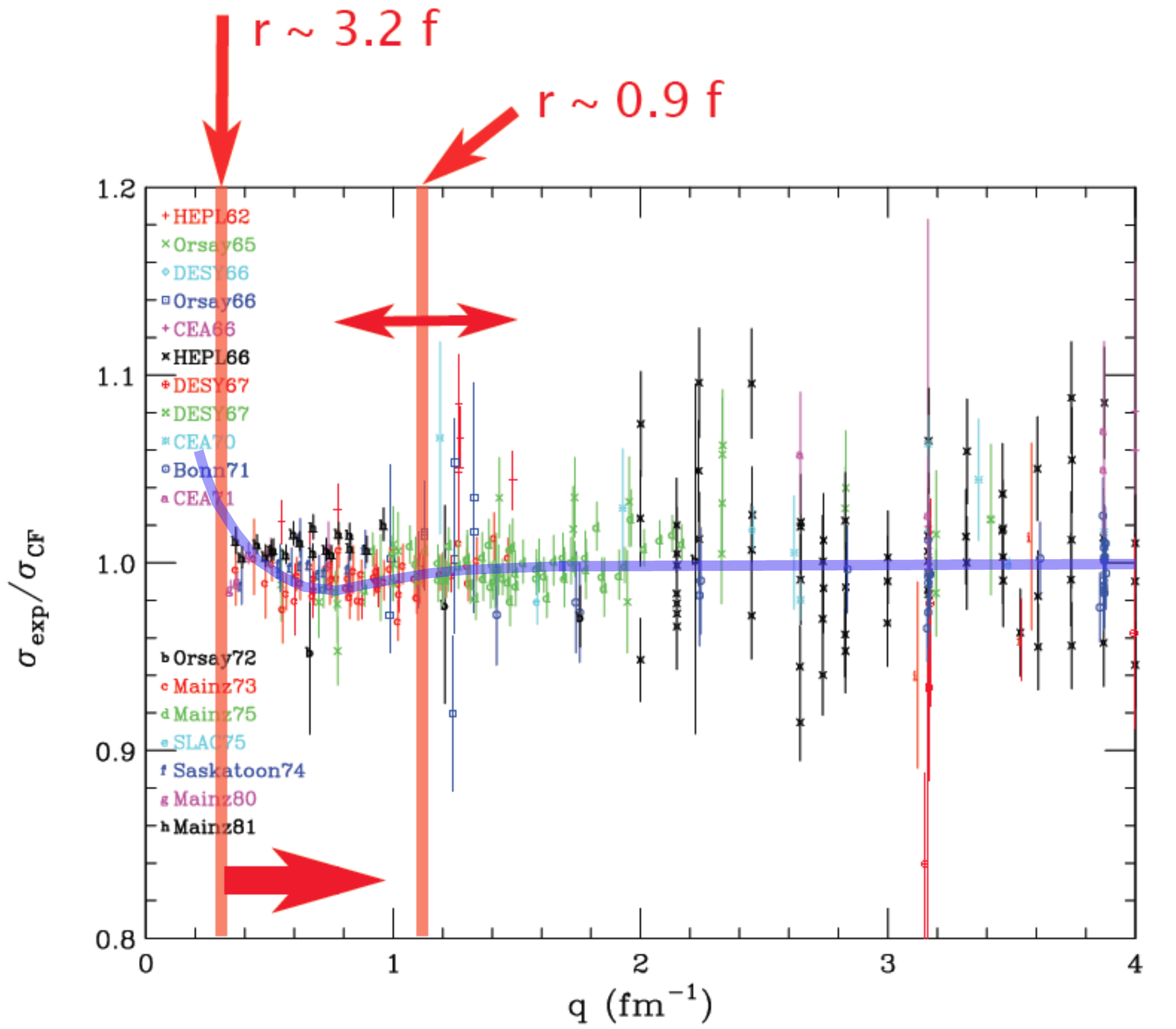}
\caption{
Low-$|{\mathbf q}|$ data, compiled and analized by Sick \cite{Sick}. 
The lines are my addition. Only $r\!\sim\!0.9$ f is bracketed
by the data, which are all to one side of $r\!\sim\!3.2$ f.
The continuous curve --drawn assuming that the absolute data
normalization is not sacred-- illustrates a possible shape
whose corresponding $\rho_p(r)$ would have a conventional
$\langle r_p^2 \rangle$ and a ``large" $\langle r_p^3 \rangle_{(2)}$.
  \label{fig:Sick}}
\end{figure}

\section{Discussion}

Very soon after ``v2" of this paper appeared in arXiv, a preprint by 
Cl$\ddot{\rm o}$et and Miller was posted \cite{CM}.
These authors find it impossible to extract a result as
large as that of Eq.~(\ref{finalresult}) from $ep$ scattering data.

A crucial problem in this connection
was adroitly emphasized by Sick \cite{Sick}. 
It is very difficult to extract reliable information on $\rho(r)$ 
from its Fourier
transform, $G_E({\mathbf q})^2$.
The radius of convergence
of the expansion in $r$ from which one extracts $\langle r_p^2\rangle$ 
is so small, that one must use simulations 
and a continued-fraction expansion 
to obtain a stable, numerically-meaningful
result not contaminated, for instance, by the term in $\langle r_p^4\rangle$.
Clearly, if extricating $\langle r_p^2\rangle$ is delicate, the more so it is
to infer $\langle r_p^3 \rangle_{(2)}$. 

A problem in extracting  $\langle r_p^3 \rangle_{(2)}$
from $ep$ data is illustrated in Fig.~\ref{fig:Sick}, borrowed
from \cite{Sick}. The data amply bracket a domain around
$r\!\sim \! 0.9$ f, required to measure an $\langle r_p^2\rangle^{1/2}$ of this
order. Contrariwise, these data are all to one side of $r\!\sim\! 3.2$ f, the
$\langle r_p^3 \rangle^{1/3}_{(2)}$ scale
in  Eq.~(\ref{finalresult}). A result based on them has to be
an extrapolation of data with a large spread and a poor
$\chi^2$ per degree of freedom. 

The absolute normalization of the data at small ${\mathbf q}^2$
(and their always disdained systematic errors) are a notorious hurdle \cite{Sick}.
If one tolerates a few \% uncertainty, the actual $G_E({\mathbf q})$
may correspond to the shape shown in Fig.~\ref{fig:Sick}, reminiscent
of the $\pi\pi$ contribution to $F_{1,2}$ calculated in \cite{DispersionRelations}.
The Fourier
transform of this shape would have the customary $\langle r_p^2 \rangle$ but a
``surplisingly" large $\langle r_p^3 \rangle_{(2)}$.

Theoretical estimates of $\langle r_p^3 \rangle_{(2)}$ would be interesting.
In their study \cite{DispersionRelations}, Belushkin et al.~find that
the $2\pi$ continuum is a very significant contribution, rising dramatically
below its scale, $|{\mathbf q}|\!=\!2\,m_\pi\!\sim\! 0.71$ f$^{-1}$, and requiring
for its  parametrization products of up to three poles.

\subsection*{Acknowledgments}
I am grateful to Alberto Galindo, Shelly Glashow, Hans-Werner Hammer 
and Ulf Meissner
for discussions and to Eduardo de Rafael for good quizzes and for directing me to an error
in my original Eq.~(\ref{DeltaE3}).
I am also indebted to York Schroeder and  to Ian
Cl$\ddot{\rm o}$et and Gerald Miller for having independently found a mistake 
in my original explicit $\langle r^3\rangle_{(2)}$ expressions.
York's interest and remarks were of great help for me.


\begin{thebibliography}{999}


\bibitem{Pohl}
R.~Pohl,  {\it et al.} Nature {\bf 466} 213 (2010) 213.
\bibitem{Flowers}
J.~Flowers ``News and Views" on \cite{Pohl},
{\it ibid.} 195.
\bibitem{PohlAddendum}
Supplementary material to \cite{Pohl}.
\bibitem{DispersionRelations}
M.~A.~Belushkin, H.~W.~Hammer \& U.~G.~Meissner, Phys. Rev. {\bf C75} (2007)  035202.
\bibitem{PionCloud}
N.~Bohr, Faraday Lecture, London, May 8th (1930).
\bibitem{HTheory}
M.~I.~Eides, H.~Grotch \& V.~A.~Shelyuto,  Springer
Tracts in Mod.~Phys.  {\bf 222}  (Springer, Berlin Heidelberg,
2007).
S.~G.~Karshenboim, Phys. Rep. {\bf 422} (2005) 1.
\bibitem{Hexps}
M.~Niering,   Phys. Rev. Lett. {\bf 84} (2000) 5496. 
B.~de Beauvoir,   Eur. Phys. J. {\bf D12} (2000) 61.
C.~Schwob Phys. Rev. Lett. {\bf 82} (1999) 4960.
\bibitem{CODATA}
P.~J.~Mohr, B.~N.~Taylor  \& D.~B.~Newell,  Rev. Mod. Phys. {\bf 80} (2008) 633.
\bibitem{Sick}
I.~Sick,  Phys. Lett. {\bf B576} (2003) 62.
\bibitem{Sick2}
P.~G.~Blunden, \& I.~Sick,  Phys. Rev. {\bf C72}  (2005) 057601.
\bibitem{LymanTH}
S.~G.~Karshenboim,   Phys. Rep. {\bf 422} (2005) 1.
K.~Pachucki,
Phys. Rev. {\bf A60} (1999) 3593.
 E.~Borie,  Phys. Rev. {\bf A71}  (2005) 032508.
A.~P.~Martynenko,  Phys. Rev. {\bf A71} (2005) 022506. 
A.~P.~Martynenko, 
 Phys. At. Nucl. {\bf 71} (2008) 125.
 K.~Pachucki,\& U.~D.~Jentschura,   Phys. Rev. Lett. {\bf 91}  (2003) 113005.
\bibitem{FriarSick}
J.~L.~Friar \& I.~Sick  Phys. Rev. {\bf A72}  (2005) 040502(R).
\bibitem{Balmer}
J.~J.~Balmer,
Verhandlungen der Naturforschenden
Gesellschaft in Basel {\bf 7} (1885) 548.
\bibitem{CM}
I.~C.~Cl$\ddot{\rm o}$et \& G.~A.~Miller, arXiv:1008.4345v1

\end{thebibliography}
\end{document}